\documentclass[aps, prapplied, twocolumn, secnumarabic, amssymb, superscriptaddress]{revtex4-2}

\usepackage{graphicx}
\graphicspath{{Figures/}} 

\usepackage{dcolumn}
\usepackage{bm}

\usepackage[utf8]{inputenc}
\usepackage[T1]{fontenc}
\usepackage{mathptmx}

\usepackage{upgreek} 
\usepackage{xcolor} 

\begin{document}

\title[SRON Ti/Au TES Calorimeters]{Performance of the SRON Ti/Au Transition Edge Sensor X-ray Calorimeters}

\author{Martin de Wit}
 \affiliation{NWO-I/SRON Netherlands Institute for Space Research, Niels Bohrweg 4, 2333 CA Leiden, The Netherlands}
 \email{M.de.Wit@SRON.nl}
\author{Luciano Gottardi}
 \affiliation{NWO-I/SRON Netherlands Institute for Space Research, Niels Bohrweg 4, 2333 CA Leiden, The Netherlands}
\author{Kenichiro Nagayoshi}
 \affiliation{NWO-I/SRON Netherlands Institute for Space Research, Niels Bohrweg 4, 2333 CA Leiden, The Netherlands}
\author{Hiroki Akamatsu}
 \affiliation{NWO-I/SRON Netherlands Institute for Space Research, Niels Bohrweg 4, 2333 CA Leiden, The Netherlands}
\author{Marcel P. Bruijn}
 \affiliation{NWO-I/SRON Netherlands Institute for Space Research, Niels Bohrweg 4, 2333 CA Leiden, The Netherlands}
\author{Marcel L. Ridder}
 \affiliation{NWO-I/SRON Netherlands Institute for Space Research, Niels Bohrweg 4, 2333 CA Leiden, The Netherlands}
\author{Emanuele Taralli}
 \affiliation{NWO-I/SRON Netherlands Institute for Space Research, Niels Bohrweg 4, 2333 CA Leiden, The Netherlands}
\author{Davide Vaccaro}
 \affiliation{NWO-I/SRON Netherlands Institute for Space Research, Niels Bohrweg 4, 2333 CA Leiden, The Netherlands}
\author{Jian-Rong Gao}
 \affiliation{NWO-I/SRON Netherlands Institute for Space Research, Niels Bohrweg 4, 2333 CA Leiden, The Netherlands}
 \affiliation{Optics Research Group, Department of Imaging Physics, Delft University of Technology, Van der Waalsweg 8, 2628 CH, Delft, The Netherlands} 
\author{Jan-Willem A. den Herder}
 \affiliation{NWO-I/SRON Netherlands Institute for Space Research, Niels Bohrweg 4, 2333 CA Leiden, The Netherlands}
 \affiliation{Anton Pannekoek Institute, University of Amsterdam, Science Park 904, 1098 XH Amsterdam, the Netherlands} 

\date{\today}

\begin{abstract}
In the early 2030s, ESAs new X-ray observatory, Athena, is scheduled to be launched. It will carry two main instruments, one of which is the X-ray Integral Field Unit (X-IFU), an X-ray imaging spectrometer, which will consist of an array of several thousand transition-edge sensors (TESs) with a proposed energy resolution of 2.5~eV for photon energies up to 7~keV. At SRON we develop the backup TES array based on Ti/Au bilayers with a transition temperature just below 100~mK. In this contribution we will give a broad overview of the properties and capabilities of these state-of-the-art detectors. Over the years we have fabricated and studied a large number of detectors with various geometries, providing us with a good understanding of how to precisely control the properties of our detectors. We are able to accurately vary the most important detector properties, such as the normal resistance, thermal conductance and critical temperature. This allows us to finely tune our detectors to meet the demands of various applications. The detectors have demonstrated excellent energy resolutions of below 1.8~eV for 5.9~keV X-rays. By tuning the properties of the devices, they can be optimally matched to various read-out schemes using both AC and DC biasing. The next step is to increase the size of our TES arrays from our current kilo-pixel arrays towards the full-sized array for X-IFU. 
\end{abstract}

\keywords{Transition Edge Sensors, X-ray spectroscopy, X-IFU, Athena}

\maketitle

\section{Introduction} \label{sec:intro}

The X-ray Integral Field Unit (X-IFU) is one of two instruments planned to be onboard of Athena, ESA's new X-ray observatory planned to be launched in the 2030s \cite{Barret2020}. X-IFU is a cryogenics imaging spectrometer based on an array of several thousand transition-edge sensors (TESs). These TESs are expected to have an energy resolution better than 2.5~eV for photon energies up to 7~keV, and be sensitive to photons in a wide energy band between 0.2 and 12~keV \cite{Barret2016}. At SRON we are developing the backup detector array for X-IFU. Other potential applications of this technology are for HUBS \cite{Cui2020}, LiteBIRD \cite{Hazumi2019}, or laboratory-based applications such as material analysis, laboratory astrophysics, hot plasma physics, and diagnostic for future fusion reactors. \cite{Hoover2009, Ullom2015, Eckart2021}.

\begin{figure*}
	\begin{center}
   		\includegraphics[width=0.9\linewidth]{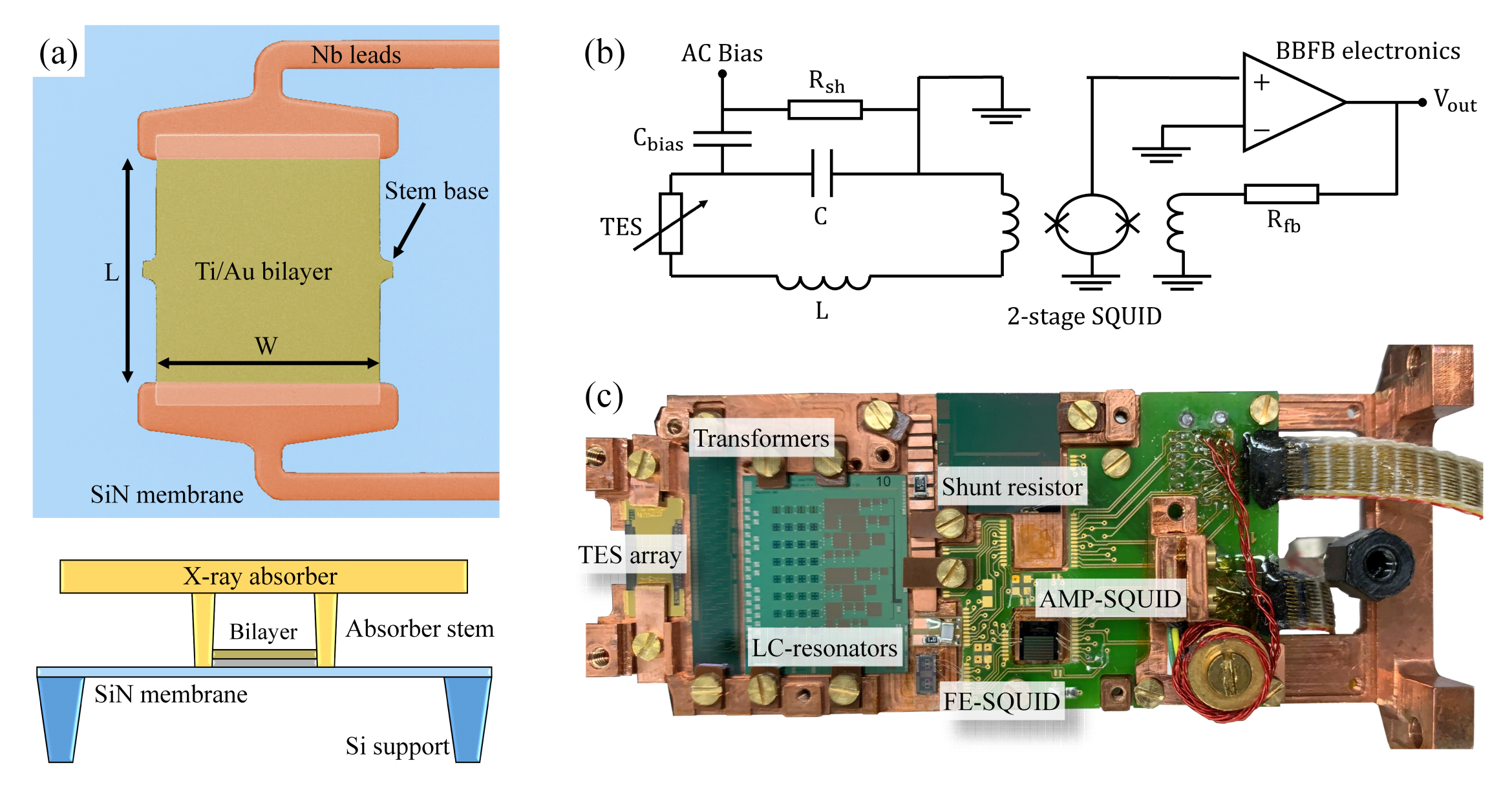}
	\end{center}
	\caption{\label{fig:Setup}(a) Colorized scanning electron microscope image of top view of a TES detector without absorber (top left), and a schematic side-view of the TES (bottom left). Not shown are four additional stems supporting the absorber on the membrane. (b) Schematic circuit of the FDM readout, with all mayor components labeled. (c) Picture of the XFDM-Probe setup used to acquire the data presented in this work, with the main components labeled. The transformer chip is optional and is used to better match the detectors to the SQUID readout. In this paper we have used either a DC-interconnection or a transformer with a 1:2 transformer turn ratio.}
\end{figure*} 

Several excellent review papers have been published and explain TESs and their applications in great detail \cite{Irwin2005, Ullom2015, Gottardi2021a}. A TES can be summarized as being composed of three vital components. The first is a sensitive thermometer, which in the case of a TES is made of a superconducting bilayer with a very sharp superconducting-to-normal state transition. When the TES is operated such that it is biased within the superconducting transition, minor changes in temperature can be detected with great accuracy. In order to absorb the energy of incoming photons that cause these temperature changes the TES is coupled to an absorber, the second main component. For X-rays, this absorber is a micro-fabricated slab of metal with a low heat capacity and a high photon stopping power. Together this ensures a high photon counting efficiency combined with a large temperature change when a photon is absorbed. Finally, to isolate the TES from the thermal surroundings the superconducting bilayer is fabricated on top of a silicon nitride membrane. A top view and schematic cross-section of a typical TES is visible in Fig. \ref{fig:Setup}(a). A more detailed explanation will follow in Sec. \ref{sec:TES}.

To readout a large number of TESs, as is required for X-IFU, it is necessary that an appropriate multiplexing readout is developed. Several readout technologies are available, of which we will name a few of the most mature here. In Time Division Multiplexing (TDM) \cite{Doriese2016} and Code Division Multiplexing \cite{Morgan2016} all detectors are DC biased and signals are readout sequentially by in turn switching on and off the various detector channels. Frequency Domain Multiplexing (FDM) \cite{Akamatsu2020, Akamatsu2021} is fundamentally different as the detectors are AC biased and signals are readout simultaneously at different MHz frequencies. Finally Microwave SQUID multiplexing \cite{Nakashima2020} can be considered a hybrid in the sense that the detectors are DC biased but the signals are modulated and measured at GHz frequencies. At SRON, we have historically focused on the development of TES detectors intended for FDM which favors detectors with a high resistance ($\gtrsim 100$~m$\Omega$) \cite{Taralli2019, Wit2020}. However, due to the recent decision to use TDM in the X-IFU instrument, we have also started to develop detector more suitable for TDM which performs better using low resistance devices \cite{Taralli2022}.

The necessity to fabricate devices for these different readout techniques has driven us to fabricate devices with a wide variety of properties but all with a very good performance. We will discuss how the design of the TESs can be used to control the main detector properties, e.g. the normal resistance $R_n$, the thermal conductance $G_b$ and critical temperature $T_C$, and the susceptibility to magnetic fields. The capability to control and change these parameters is of vital importance for the successful operation of both the individual detectors, as well as their multiplexed readout. By careful tuning one can control the speed of the detectors, which in turn determines the maximum count rate and cross talk, as well as the final energy resolution, which to first order scales with $T_C^{3/2}$. The pixel design is a compromise between the different properties such that the final detector is optimally suited to the desired application. For simplicity, the detectors that were used for this study use pure gold (Au) X-ray absorbers. New absorbers that also contain a layer of bismuth (Bi) with a higher quantum efficiency have been developed, but details of their fabrication and performance will be reported elsewhere.

\section{TES detectors and FDM readout} \label{sec:TES}

Over the years a wide variety of TES detectors has been fabricated in order to match different specifications and readout technologies. The basic structure of our detectors is shown in Fig. \ref{fig:Setup}(a). The heart of the detectors is a Ti/Au superconducting bilayer that is used as a highly sensitive thermometer. The thickness of the two layers can be used to vary the square resistance and critical temperature of the bilayer. The high temperature sensitivity of this type of detectors arises from the steep transition between the superconducting and normal state of the bilayer in which a small change in temperature causes a large change in the resistance of the detector. In order to decouple the TES from the thermal bath the bilayers are grown on a 0.5~$\upmu$m silicon-nitride membrane. The energy of incoming photons is measured by coupling the TESs to a X-ray absorber via two central stems placed on the ears visible in the yellow bilayer of Fig. \ref{fig:Setup}(a). The absorber is not visible in the top figure of \ref{fig:Setup}(a), but it is included in the schematic side view in the bottom figure. An additional set of four stems are placed on the membrane to support the weight of the absorber. The bilayer is connected to the rest of the readout circuit using niobium leads with a critical temperature of 9~K.

All data presented in this paper was measured using an alternating bias current to keep the detectors within the superconducting transition. This was done by using one of our FDM setups in single pixels mode. The schematic bias circuit of a single pixel in FDM is shown in Fig. \ref{fig:Setup}(b). Each TES is biased using a carrier signal with a frequency between 1 and 5 MHz. Each TES is coupled to a high-Q superconducting LC resonator to define a specific resonance frequency used for biasing. We use a coil inductance of 1~$\upmu$H and vary the capacitance to get the desired bias frequency for each detector. The TES is operated in a stiff voltage bias using a shunt resistor $R_\mathrm{sh}$, which together with the bias capacitor $C_\mathrm{bias}$ (chosen such that $C/C_\mathrm{bias} = 25$) leads to an effective shunt resistance of 1.2~m$\Omega$. The current that runs through the TES is measured using a two-stage SQUID amplifier. Base-band feedback (BBFB) is used to increase the linearity and dynamic range. More information on the FDM readout scheme is given in Akamatsu \textit{et al.}\cite{Akamatsu2021}. A picture of the practical implementation of this circuit is shown in Fig. \ref{fig:Setup}(c) with the main components labeled. The setup shown in this figure is mounted at the mixing chamber stage of a dilution refrigerator and is operated at a temperature of 50~mK. A Helmholtz coil is placed over the TES array in order to apply a uniform magnetic field perpendicular to the TESs. Magnetic shielding of the setup is achieved using a lead and cryoperm shield at the cold stage, and a mu-metal shield around the outside of the cryostat. The setup is discussed in more detail elsewhere \cite{Wit2020}.

\section{Impact of bilayer design on detector properties} \label{sec:Impact_Design}

\begin{figure*}
	\begin{center}
   		\includegraphics[width=0.9\textwidth]{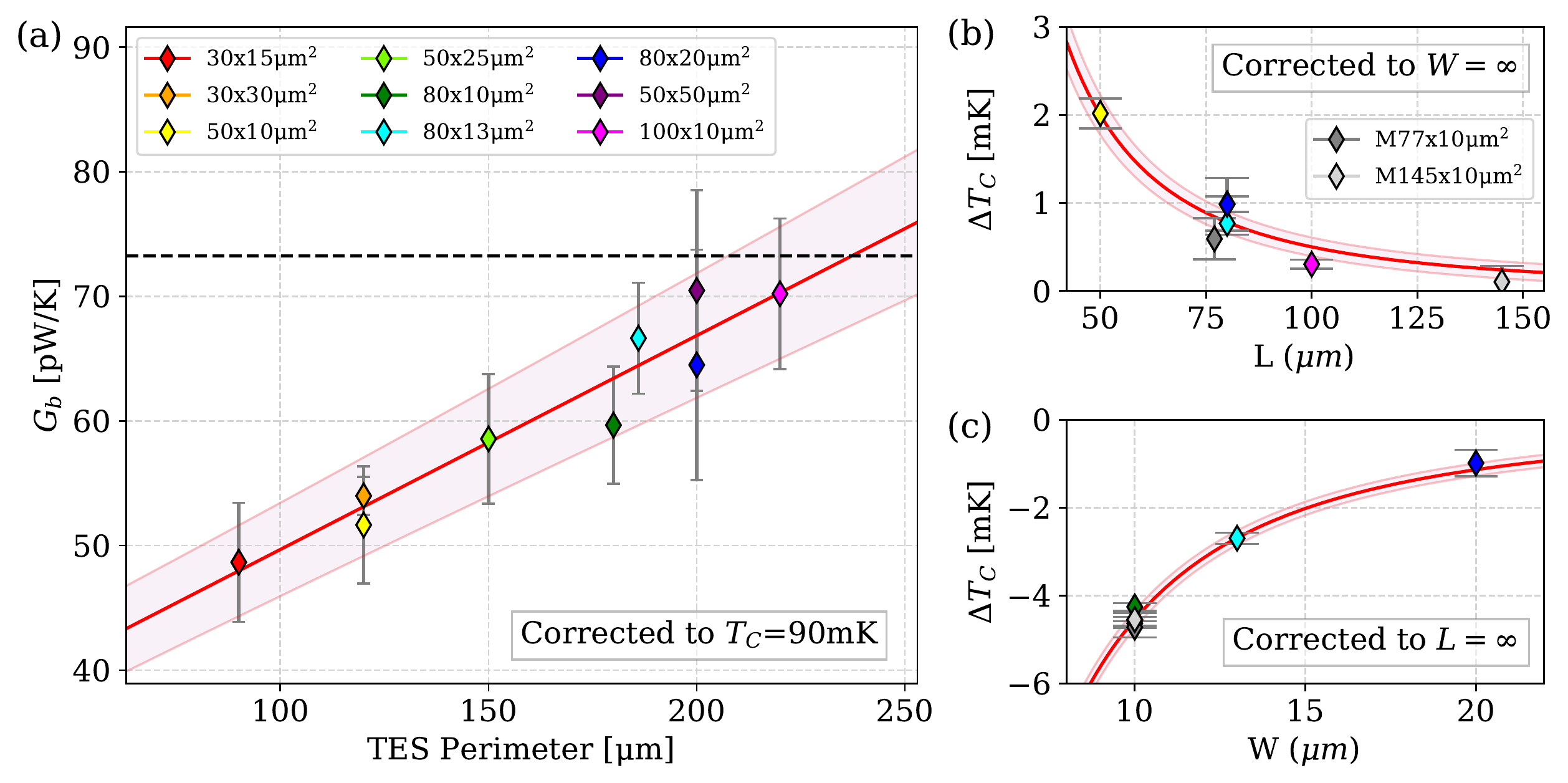}
	\end{center}
	\caption{\label{fig:G-Tc}(a) Thermal conductance between the bilayer and the bath as a function of the bilayer perimeter ($2W + 2L$). The solid red line is a linear fit following a radiative transport model (filled area indicates 1~$\sigma$ uncertainty). The black dashed line shows the current baseline for X-IFU. (b) Shift of $T_{C,\mathrm{i}}$ as a function of the bilayer length after correction for the LaiPE ($W = \infty$). (c) Shift of $T_{C,\mathrm{i}}$ as a function of the bilayer width after correction for the LoPE ($L = \infty$). The solid red lines are fits to quadratic dependencies on the length and width (filled area indicates 1~$\sigma$ uncertainty).}
\end{figure*} 

In this section we discuss the impact of the TES design on a number of critical TES properties. In particular, we cover the normal resistance $R_n$, the thermal conductance between the bilayer and the bath $G_b$, the critical temperature $T_C$, and the sensitivity to magnetic fields. To map these properties we have measured a number of TESs with varying dimension of the bilayer, identified using the naming convention $L \times W\upmu$m$^2$ (as defined in Fig. \ref{fig:Setup}(a)). These devices are rectangular, except when the name is preceded by an ``M", which indicates that this design is a meandering geometry. These meandering devices are intended to allow for longer bilayers to fit within the limited space of the silicon-nitride membrane. The devices that are compared here come from a single wafer but from different arrays which are spatially seperated on the wafer. The bilayers have an Ti/Au thickness of 35/200~nm, leading to a critical temperature of $\sim$~90~mK and a measured square resistance of 25.8~m$\Omega$/$\square$ (unless specified otherwise). The bilayers are coupled to a 240$\times$240~$\upmu$m$^2$ Au X-ray absorber with a thickness of 2.35~$\upmu$m. With a pixel pitch of 250~$\upmu$m this leaves a gap of 10~$\upmu$m between two adjacent absorbers. 

The most straight-forward way in which $R_n$ of the devices can be changed is by adjusting the thickness of the layers of the bilayer. However, the fabrication of the bilayer is a highly tuned process where reproducible is of the utmost importance. Therefore, a more practical approach to control the normal resistance is to appropriately choose the bilayer aspect ratio (AR = $L$:$W$). In this section we present data for devices with aspect ratios ranging from 1:1 up to 14.5:1, meaning the normal resistance ranges from 26~m$\Omega$ up to 374~m$\Omega$. With the AR fixed by the requirements for the desired $R_n$ for the readout scheme, one is still free to change the absolute size of the bilayer. The impact of changing the width and length of the detectors on the thermal properties is shown in Fig. \ref{fig:G-Tc}, with (a) showing the thermal conductance between the bilayer and the bath and (b, c) the critical temperature. $T_C$ and $G_b$ are measured using the standard approach of measuring IV curves as a function of bath temperature, and fitting the extracted TES power versus temperature curve using a power law model \cite{Lindeman2008}.

The thermal conductance, evaluated at $T_C$, is corrected to a $T_C$ of 90~mK to enable the direct comparison of different devices. The dominant process that determines the thermal conductance in this type of devices is two-dimensional radiative transport in the silicon nitride membrane \cite{Hoevers2005,Hays-Wehle2016}. This type of transport is characterized with a linear dependence between $G_b$ and the radiating perimeter. The measured $G_b$ plotted as a function of the perimeter of the bilayer, as shown in Fig. \ref{fig:G-Tc}(a), verifies that the current results are consistent with this interpretation. We obtain a proportionality constant between the perimeter and $G_b$ of 0.17~pW/K/$\upmu$m, and an offset of approximately 30 pW/K, which is attributed to parallel power radiation from the 4 supporting stems of the absorber \cite{Taralli2021}. Thus, by changing the absolute size of the bilayer whilst maintaining a constant AR, it is possible to tune $G_b$ without changing $R_n$, as demonstrated by comparing i.e. the 30$\times$15~$\upmu$m$^2$ (red) and the 50$\times$25~$\upmu$m$^2$ (light green) devices.

Note that the meandering devices are not included in Fig. \ref{fig:G-Tc}(a). The measured $G_b$ of these devices falls below the expected trend from the rectangular devices, indicating that a simple radiative model is insufficient to fully explain the data. We are working under the assumption that this is due to partial reabsorbtion of the radiated heat by opposing sections of the bilayer.

The intrinsic critical temperature of a TES ($T_{C,\mathrm{i}}$) is to first order determined by the properties of the bilayer such as the choice of material, the thickness of the layers, and the quality of the interface\cite{Martinis2000}. However, real TESs can have an effective $T_C$ different from $T_{C,\mathrm{i}}$, depending on both the width and the length of the devices. This dependence is due to a combination of the longitudinal proximity effect (LoPE) from the higher Tc Nb leads connected to the bilayer and the lateral inverse proximity effect (LaiPE) from the edges of the bilayer \cite{Sadleir2010, Sadleir2011}. The geometry dependent shift of the critical temperature $\Delta T_C = T_C - T_{C,\mathrm{i}}$ is shown in the right-hand-side plots of Fig. \ref{fig:G-Tc}, where we plot separately the impact of the length (a) and width (b) and the bilayer. The effective critical temperature in the presence of both these effect can be empirically described by
\begin{equation}
	T_C = T_{C,\mathrm{i}} + C_{W}/W^2 + C_{L}/L^2,
\end{equation}
with $C_{W}$ and $C_{L}$ the proportionality constants related to the LaiPE and LoPE. In this equation we assume that the two proximity effects are independent.

In order to independently evaluate the impact of the LoPE and the LaiPE, the data shown in Figs. \ref{fig:G-Tc}(b) and (c) has been corrected to infinite width and length, respectively, following the method explained elsewhere \cite{Wit2020}. In both figures $\Delta T_C$ is calculated by subtracting the same $T_{C,\mathrm{i}} = 86.7$~mK. The extracted proportionality constants are $C_{W} = -454 \pm 15$~mK\,$\upmu$m$^2$ and $C_{L} = 4997 \pm 518$~mK\,$\upmu$m$^2$. The large value for $C_{L}$ reflects the large difference between the $T_C$ of the bilayer and that of the niobium leads. The value of $C_{W}$ is seen to vary between detector arrays due to variations in the edge conditions resulting from the fabrication. From the figures it is clear that variations in the TES design can easily shift the critical temperature by several mK, indicating the importance of considering these effects properly when a very accurate value for $T_C$ has to be attained.

Apart from the thermal properties, another important aspect of the TES is the susceptibility to external magnetic fields. It has long been known that TESs act as weak superconducting links due to the coupling between the bilayer and the high $T_C$ superconducting leads \cite{Sadleir2010, Smith2013a, Gottardi2014}. One of the consequences of this weak-link behaviour is that the TES (critical) current exhibits oscillations with an applied magnetic field perpendicular to the TES array. For an ideal weak link measured at temperatures above $T_{C,\mathrm{i}}$, these oscillations are well-described using a Fraunhofer pattern with period $\Delta B = \Phi_0/A_\mathrm{wl}$, where $\Phi_0 = 2\cdot10^{-15}$~Wb is the magnetic flux quantum and $A_\mathrm{wl}$ is the effective area of the bilayer that acts as a weak-link. The shape of the oscillatory pattern becomes more complicated when a non-uniform current distribution is present in the bilayer \cite{Barone1982}, or when the $I_C(B)$ is measured at temperatures well below $T_{C,\mathrm{i}}$, in which case screening of the applied field due to the Meissner effect cannot be neglected, effectively convolving the Fraunhofer-like pattern with a triangular field dependence.

\begin{figure*}  
	\begin{center}
   		\includegraphics[width=0.9\textwidth]{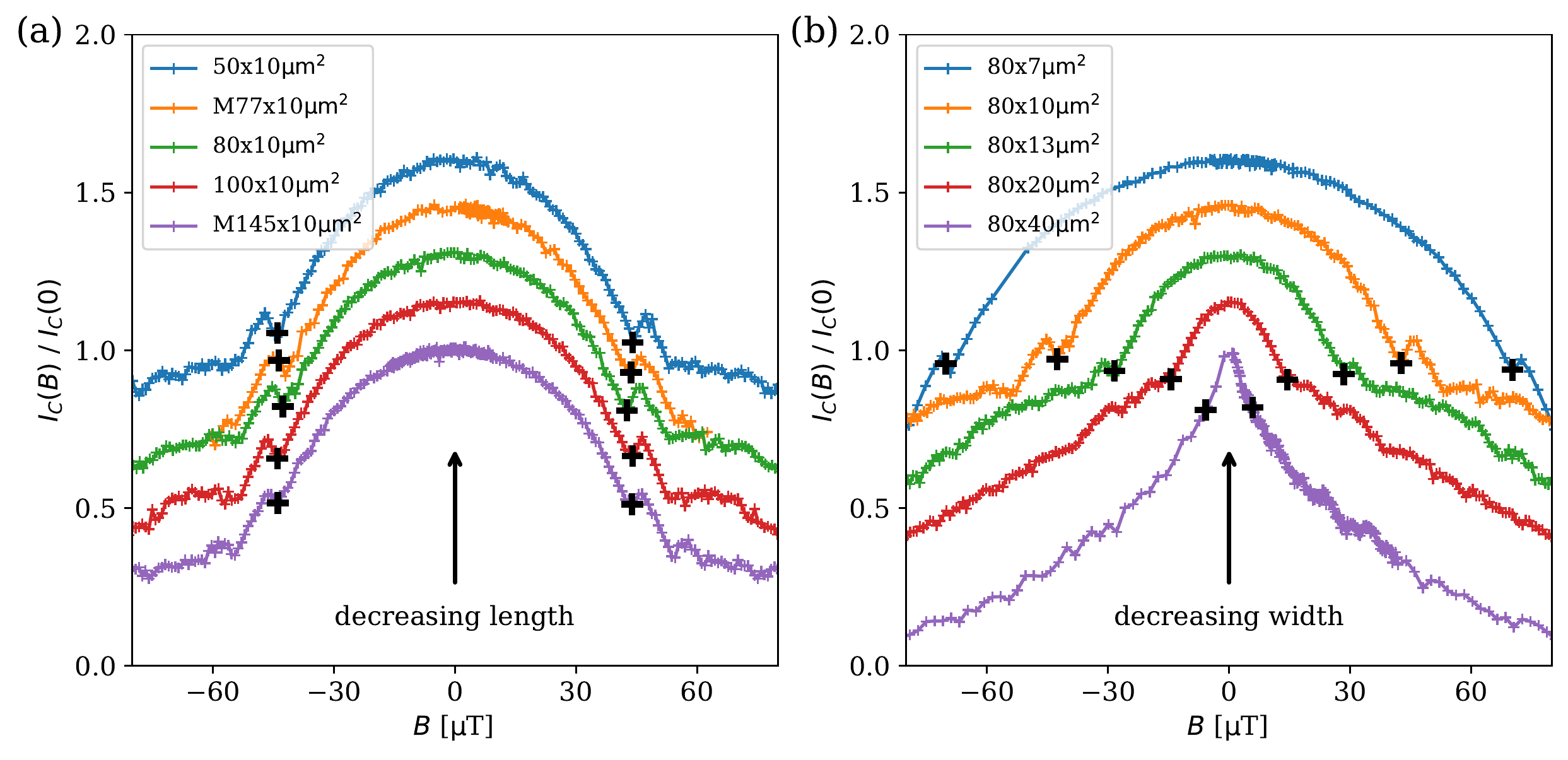}
	\end{center}
	\caption{\label{fig:IcB}Measured $I_C(B)$, normalized to $I_C(0)$, for a selection of different TES designs split into two sets: (a) devices of equal width with varying length, and (b) devices of equal length with varying width. The curves are artificially offset for easier comparison.}
\end{figure*} 

In order to demonstrate the magnetic field susceptibility, we have measured the critical current as a function of the perpendicular applied magnetic field for a number of different TES designs. In general, the magnetic field dependence exhibited by the $I_C(B)$ is a good reflection of other effects caused by the magnetic field, such as changes in the energy scale calibration \cite{VaccaroXXXX}. In Fig. \ref{fig:IcB} we show the measured $I_C(B)$, normalized to $I_C(0)$, for a number of different bilayer designs. In (a) we show a selection of devices of equal width ($W = 10~\upmu$m) but lengths varying from 50 to 145~$\upmu$m, while in (b) we show devices of equal length ($L = 80~\upmu$m) and widths varying between 7 and 40~$\upmu$m. The curves were measured at a bath temperature of 50 mK, well below $T_{C,\mathrm{i}}$. The noise visible in the $I_C(B)$ curves arises from the way in which the measurement is done: the TES current is increased starting from the superconducting state. When an X-rays hits the absorber with $I \sim I_C(B)$ the TES jumps to the normal state at a slightly lower current than the unperturbed $I_C$. An X-ray hits the detector at about 1 count per second, leading to some jitter in the final measured $I$ before the transition. It is clear that none of the curves resembles a clean Fraunhofer pattern with a clearly distinguishable oscillation period. This observation can be understood from the reasons stated above, namely the presence of Meissner screening and a spatially inhomogeneous current distribution. While in theory it is possible to reconstruct the current distribution from the shape of the $I_C(B)$ curve it is by no means trivial. In general, the broad central maximum and narrow suppressed side lobes are typical for a current distribution that peaks in the center of the bilayer, consistent with the expected suppression of the superconductivity at the edges of the bilayer \cite{Sadleir2011,Wit2020}.

When we take the first clearly observable local minimum next to the central peak (indicated with the black markers in Fig. \ref{fig:IcB}) as a value for the oscillation period, one clearly sees that the width of the central maximum is independent of the length of the bilayer and is only affected by the width of the bilayer. This observation is surprising when considering the weak-link framework introduced above, where one would expect the position of the first minimum to depends on the total affected area, so both on the length and the width. Additionally, the period that we observe is much larger than what one would calculate based on the area of the bilayers, which would be only a few $\upmu$T depending on which design is evaluated. These points suggest that the broad central maximum results from an effective area that is much smaller than the total area of the bilayer. In fact, when we assume that $A_\mathrm{wl} = WL_\mathrm{eff}$, we find values for $L_\mathrm{eff}$ between 4-7~$\upmu$m, independent of the length of the bilayer. In a previous study, we have shown that the effective length is not located near the center of the bilayers where the absorber stems are connected \cite{DeWit2021}, which leaves the area right next to the leads as the main suspect. A detailed explanation remains elusive. More subtle effects could be present in the $I_C(B)$ resulting from a larger portion of the bilayer, which would have a much smaller period, but these effects are obscured by the very broad central feature.  Additionally, it is possible that devices with such high AR start to migrate away for weak-link behaviour and should instead be considered in the context of phase-slip lines \cite{Likharev1979, Gottardi2021a}.

Despite this lack of a full understanding of the observed $I_C(B)$, the conclusion from Fig. \ref{fig:IcB} is that it is possible to tune the sensitivity to magnetic fields purely by changing the width of the detectors. The length of the bilayer remains a free parameter that can be used to change the normal resistance or thermal conductance to match the detector requirements.

\section{Noise-Equivalent Power and Energy Resolution}

Arguably the most important figure of merit for TES-based X-ray calorimeters is the achievable energy resolution $\Delta E$. A better energy resolution, or more generally speaking a higher resolving power $E/\Delta E$, means the instrument has a higher capability of resolving closely-spaced lines and measuring lineshifts. In this section we demonstrate the energy resolution in two ways, first by using the Noise-Equivalent Power ($\mathrm{NEP}$) and then by acquiring high count energy spectra of the Mn-K$\alpha$ complex.

\begin{figure*}  
	\begin{center}
   		\includegraphics[width=0.9\textwidth]{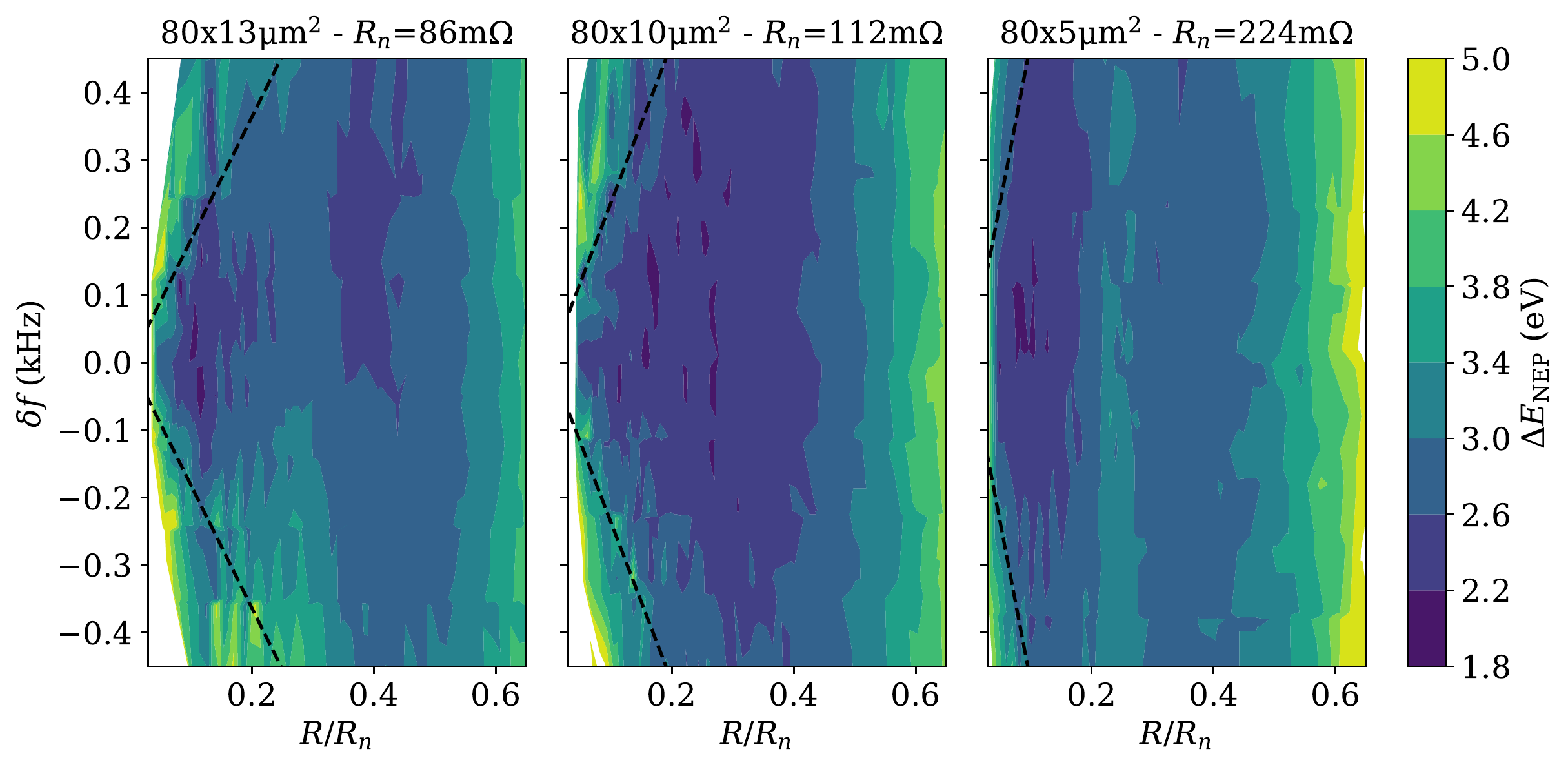}
	\end{center}
	\caption{\label{fig:NEP}Expected energy resolution from the integrated $\mathrm{NEP}$ for three pixel designs: 80$\times$13~$\upmu$m$^2$, 80$\times$10~$\upmu$m$^2$, and 80$\times$5~$\upmu$m$^2$. The detectors have a square resistance of 14~m$\Omega$/$\square$. All figures use the same color scale. The black dashed line shows the line $\delta f = R/(2 \pi L)$ with constant $L$ for all figures.}
\end{figure*} 

The $\mathrm{NEP}(f)^2 = \langle |N(f)|^2  \rangle / |S(f)|^2$  is defined as the ratio between the averaged noise power spectral density and the squared detector responsivity. From the $\mathrm{NEP}$ one can then calculate the expected energy resolution using
\begin{equation}
	\Delta E_{\mathrm{NEP}} = 2.355 \sqrt{ \frac{1}{ \int_0^\infty \frac{4}{\mathrm{NEP}(f)^2} df} }.
\end{equation}
The advantage of using the $\mathrm{NEP}$ is that the acquisition is relatively quick, which allows us to scan a large part of the parameter space. However, the $\mathrm{NEP}$ assumes that the noise is stationary during a pulse which is only true in the small signal limit. It does not account for the non-linearities in the detector for large signals such as high energy photon hits. In these events there is potentially a large excursion along the TES $R(T,I)$ curve where the magnitude of the various noise contributions can change significantly. However, for the devices reported here we have empirically found that the $\mathrm{NEP}$ is still a good predictor of the real energy resolution.

In Fig. \ref{fig:NEP} we show three heat maps containing the energy resolution extracted from the integrated $\mathrm{NEP}$ measured as a function of the bias point $R/R_n$ and the detuning $\delta f$ from the optimal bias frequency defined by the LC resonator. From left to right we present the data for three pixels with dimensions 80$\times$13~$\upmu$m$^2$, 80$\times$10~$\upmu$m$^2$, and 80$\times$5~$\upmu$m$^2$ with corresponding $R_n$ values of 86, 112, and 224~m$\Omega$, respectively. From the figure one can find the optimal bias point for each of these detectors. We see that when the TES normal resistance is increased (meaning a larger $R$ at a given $R/R_n$ fraction), a good energy resolution can be achieved for larger $\delta f$. The degradation in the energy resolution is the result of an additional impedance introduced when the bias frequency is detuned from the resonance frequency of the LC-resonator. The correlation between $\Delta E_{\mathrm{NEP}}$ and this impedance is illustrated in Fig. \ref{fig:NEP} with the black dashed lines which are given by $\delta f = R/(2 \pi L)$, a measure for the full-width-at-half-maximum (FWHM) of the LC-resonator assuming the TES resistance is the dominating contribution to the dissipation in the resonator. The additional detuning impedance increases the effective shunt resistance $R_\mathrm{sh}$. This leads to a corresponding decrease of the ratio between the TES resistance and the shunt resistance. This degrades the quality of the voltage bias and decreases the loopgain of the electro-thermal feedback \cite{Iyomoto2008a, Miniussi2020}. A detector with a higher $R$ can be used to mitigate the impact of the detuning impedance. On the other hand, the higher $R$ means that the ratio of the current in the TES and the SQUID noise (in our system $\sim$~5~pA/$\sqrt{\mathrm{Hz}}$) goes down. So the additional tolerance for the applied bias frequency when using high $R$ detectors might come at the expense of reduced optimal range in the bias point $R/R_n$, an effect that is also visible in the high $R/R_n$ data presented in Fig. \ref{fig:NEP}. Note that a bump in the $\mathrm{NEP}$ is observed for the 80$\times$5~$\upmu$m$^2$ at $R = 0.2-0.3R_n$. The origin of the bump is as of yet unknown. One explanation is that this can be caused by an altered current distribution in the bilayer due to the presence of the stems \cite{DeWit2021}. In any case this bump is not related to the effects discussed here.

\begin{figure*}  
	\begin{center}
   		\includegraphics[width=0.9\textwidth]{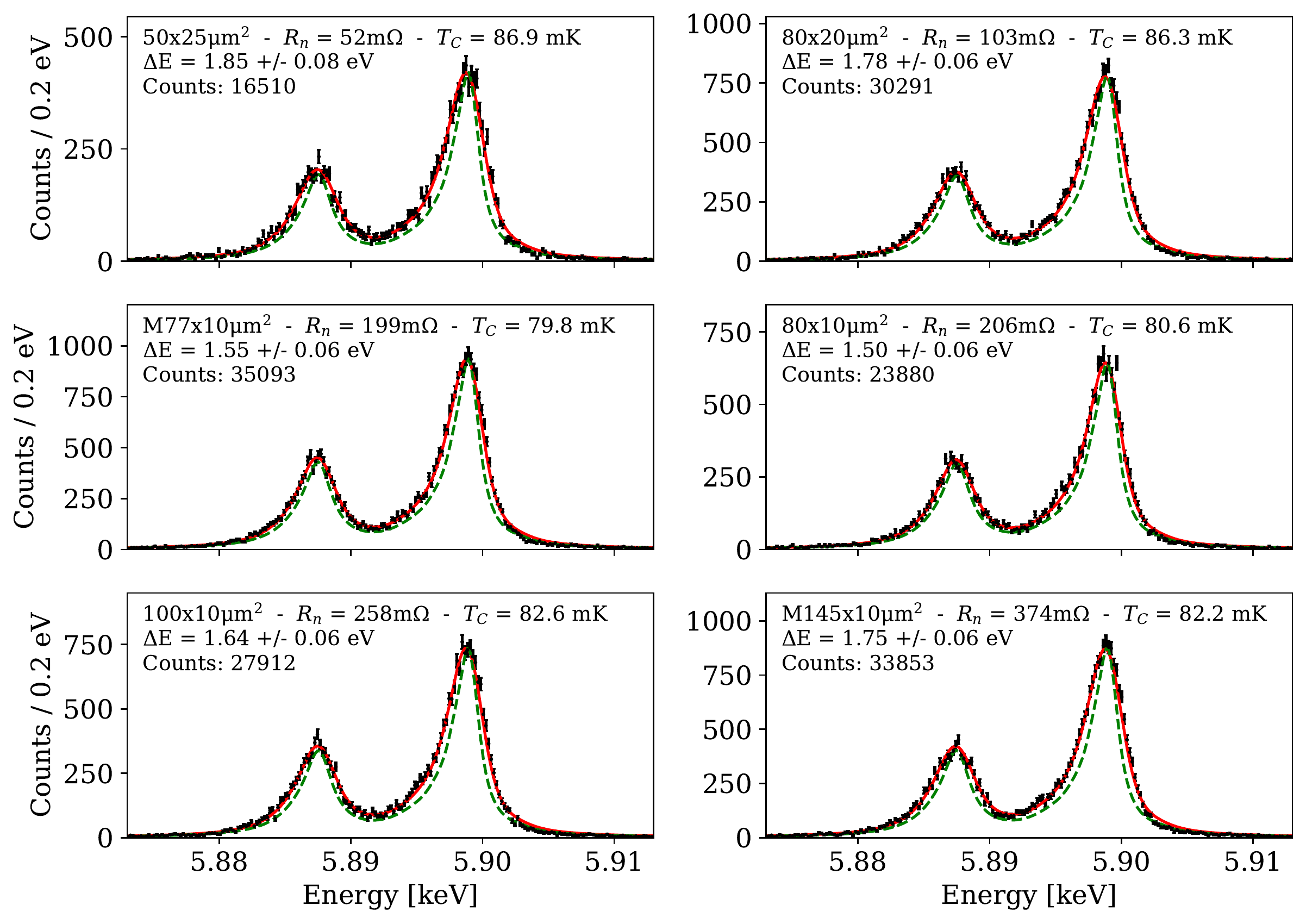}
	\end{center}
	\caption{\label{fig:Xrays}Example energy spectra measured using four different TES designs: (top left) 50$\times$25~$\upmu$m$^2$, (top right) 80$\times$20~$\upmu$m$^2$, (middle left) M77$\times$10~$\upmu$m$^2$, (middle right) 80$\times$10~$\upmu$m$^2$ (bottom left) 100$\times$10~$\upmu$m$^2$, and (bottom right) M145$\times$10~$\upmu$m$^2$. The detectors have a square resistance of 26~m$\Omega$/$\square$. General pixel properties and the achieved resolution at 5.9~keV are included at the top left of each figure. The solid red line are the best fits to the Gaussian model, while the dashed green lines show the natural line shape of the Mn-K$\alpha$ complex.}
\end{figure*} 

As stated above, while the energy resolution extracted from the integrated $\mathrm{NEP}$ is a reasonable predictor for the large-signal energy resolution, a discrepancy between the $\Delta E_{\mathrm{NEP}}$ and the X-ray energy resolution can be present depending on the inductance of the circuit $L$. Therefore the real demonstration of the resolving capabilities comes from  measuring energy spectra. A standard characterization of the detectors involves measuring the energy resolution by exposing the TES array to a $^{55}$Fe source, providing Mn-K$\alpha$ X-rays at an energy of approximately 5.9~keV. Typical count rates for this characterization are approximately 1-1.5 counts per second per pixel. The energy of each pulse is determined using the optimal filter method \cite{Szymkowiak1993}. The energy resolution is extracted from the resulting spectra by convolving the natural line shapes \cite{Holzer1997, Eckart2016} with a Gaussian instrument response. Resulting energy spectra for four example detectors are shown in Fig. \ref{fig:Xrays}.

As can be seen all presented devices achieve excellent energy resolutions well below 2~eV with $E/\Delta E > 3000$. Note that this resolution is achieved for devices with a very large range of dimensions and normal resistance, showing the flexibility of the TESs to be adjusted to optimally match the desired application. The energy resolution for the 50$\times$25~$\upmu$m$^2$ is slightly worse than that of the higher resistance devices. This originates from the mismatch between these low resistance devices and the AC-biased FDM readout \cite{Gottardi2018, Taralli2022}, as well as a too small ratio between the effective shunt resistance and the TES resistance in the bias point, reducing the stiffness of the voltage bias. This degradation is therefore not related to the properties of the detector.

\section{Summary}

To summarize, we have shown that we are capable of fabricating TES arrays with a large variety of properties. By adjusting easily controlled parameters such as the geometrical dimensions of the bilayer, it is possible to change important detector properties such as the normal resistance, thermal conductance, and critical temperature in a well-controlled way. Changing the width of the bilayer in particular determines to a large extend the sensitivity of TESs to external magnetic fields, while we have observed that the length plays no significant role. The possibility to reliably control the detector characteristics means that it is possible to match our detectors to various readout schemes, such as FDM, TDM, and microwave SQUID multiplexing. Energy resolutions well below 2~eV can be achieved the broad variety of devices.

Towards fabricating the backup array of X-IFU, the next step is to increase the size of the detector arrays from our current kilo-pixel arrays towards the full size arrays containing about 3000 detectors. Also a detailed study of the energy-dependent performance is planned, both using an already available modulated X-ray source \cite{Dandrea2021}, as well as a rotating target source currently under development. In parallel to the fabrication and testing of better detectors, the development of our FDM multiplexing technology is progressing steadily. This development is profiting directly from the improvements in the performance of individual detectors and has culminated in the recent demonstration of 2.2~eV at 5.9~keV with 37 pixels multiplexed in a single FDM readout chain \cite{Akamatsu2021}.

\section*{ACKNOWLEDGMENTS}       

SRON Netherlands Institute for Space Research is supported financially by NWO, the Dutch Research Council. This work was funded partly by NWO under the research programme Athena with Project No. 184.034.002 and partly by the European Space Agency (ESA) under ESA CTP Contract No. 4000130346/20/NL/BW/os. It has also received funding from the European Union’s Horizon 2020 Program under the AHEAD (Activities for the High-Energy Astrophysics Domain) project with Grant Agreement Number 654215.

\bibliography{SPIE_2022_bibliography}

\end{document}